\documentclass[prd,aps,twocolumn,floatfix]{revtex4-2}
\usepackage{graphicx,psfrag,mathrsfs}
\usepackage{slashed}
\usepackage{mathrsfs}
\usepackage{bbm}
\usepackage{amsmath,amsfonts,amssymb,amsthm}
\usepackage{hyperref}
\usepackage{url}
\usepackage{comment,cancel}
\usepackage{accents}
\usepackage{comment}
\usepackage{orcidlink}
\usepackage{accents}

\def\p{\partial}

\def\ul{\underline}
\def\non{\nonumber}

\usepackage[normalem]{ulem}
\usepackage{xcolor}

\begin{document}

\title{Strong hyperboloidal compactification for the spherical DF-GHG
  formulation of GR}

\begin{abstract}
  The use of compactified hyperboloidal coordinates for metric
  formulations of the Einstein field Equations introduces formally
  singular terms in the equations of motion whose numerical treatment
  requires care. In this paper we study a particular choice of
  constraint addition, choice of gauge and reduction fields in order
  to minimize the number of these terms in a spherically symmetric
  reduction of the Dual-Foliation Generalized Harmonic Gauge
  formulation of General Relativity. We proceed to the numerical
  implementation of a more aggressive compactification, as compared to
  our previous work. With the present setup there is a direct analogy
  with conformal compactification used in other approaches to the use
  of hyperboloidal coordinates. We present numerical results of
  constraints violating and satisfying perturbations on top of a
  Schwarzschild black hole. For small perturbations we recover the
  expected physics from linear theory, corresponding to quasi normal
  mode ringing and tail decay for a scalar field, both extracted
  directly at future null infinity from our numerical data.
\end{abstract}

\author{Christian Peterson$^1$\orcidlink{0000-0003-4842-1368}}
\author{David Hilditch$^1$\orcidlink{0000-0001-9960-5293}}

\affiliation{$^1$CENTRA, Departamento de F\'isica, Instituto Superior
  T\'ecnico IST, Universidade de Lisboa UL, Avenida Rovisco Pais 1,
  1049 Lisboa, Portugal }

\maketitle

\section{Introduction}

General Relativity (GR) is a diffeomorphism invariant theory. As such,
care is needed when making general statements derived upon a
particular choice of coordinates. In local statements of the theory
this issue can be easily dealt with by the use of tensors, which by
definition are gauge invariant quantities. On the other hand,
radiation in general spacetimes is more subtle, as a limit to
arbitrarily large distances should be performed to get the radiative
part of the fields at hand. This was understood
in~\cite{Bon62,Sac62a,NewPen62}, and later formalized by
Penrose~\cite{Pen64}, who realized that asymptotic quantities of
spacetime can be mathematically defined by using conformal
compactification, which leads to the definition of null
infinity,~$\mathscr{I}$. When we model isolated systems it is
customary to assume asymptotic flatness, which can be thought of as
those which resemble the Minkowski spacetime far away from the
sources~\cite{Wal84}. In such spacetimes the asymptotic region is
divided into past null infinity,~$\mathscr{I}^-$, spatial
infinity,~$i^0$, and future null infinity,~$\mathscr{I}^+$. In
particular,~$\mathscr{I}^+$ is the piece of the asymptotic region
where idealized observers sit and where future-directed radiation
travels to.

In this work we continue our research program on the inclusion
of~$\mathscr{I}^+$ in numerical relativity (NR) simulations. There are
several approaches to do so. A first example is the use of the
conformal Einstein field equations (CEFEs), first introduced
in~\cite{Fri81}, which evolves geometric quantities, including
curvature, in the conformally compactified spacetime. Despite the
CEFEs providing a regular and well-posed system of partial
differential equations (PDEs) in vacuum, and their use in influential
numerical work, see~\cite{FraSteThe25} for an interesting recent
example, they have not found widespread application for gravitational
wave (GW) astronomy. A second approach to include~$\mathscr{I}^+$ in
the numerical domain is to match Cauchy slices with compactified null
slices at each timestep. This idea goes by the name of
Cauchy-characteristic matching (CCM). See~\cite{Win12} for a review,
\cite{MaMoxSch23_a} for a recent numerical example
and~\cite{GiaBisHil21,GiaBisHil23,Gun24} for discussion of the PDE
properties of the Einstein equations in Bondi-gauge.

A third approach, used here, is to foliate spacetime with
hyperboloidal slices and use a metric formulation, without introducing
curvature as an evolved variable. The computation of GWs
at~$\mathscr{I}^+$ in black hole (BH) linear perturbation theory with
the use of hyperboloidal slices is now a standard technique which has
proven to be particularly useful. See, for
example,~\cite{ZenNunHus08,Mac20,MacZen25}. In the nonlinear setting
there are two approaches that lie in this category. One of them is to
prescribe an ad-hoc conformal factor analytically and evolve the
metric in the unphysical
spacetime~\cite{Zen08,VanHusHil14,VanHus17}. This approach assumes a
particular rate of divergence of the metric at~$\mathscr{I}^+$, which
is then taken into account by the conformal factor. This technique is
the closest metric formulation to the~CEFEs. The path we pursue here
is similar in essence, but with slightly different techniques. Our
method is based on the Dual-Foliation (DF) formalism, which constructs
a hyperboloidal slicing of spacetime from a Cauchy one by the use of a
height function~\cite{HilHarBug16}. We then compactify an outgoing
spatial coordinate to reach~$\mathscr{I}^+$ with a finite coordinate
distance. We also rescale each evolved field according to their
asymptotic decay, which are known from previous analytic
studies~\cite{DuaFenGasHil22a}. A strength of the DF technique is in
the ability to reach~$\mathscr{I}^+$ and maintain the hyperbolicity
properties of the formulation from which the Cauchy surfaces are
constructed, so that we can directly extend systems known to work well
in the strong-field region. In particular, we use the popular
Generalized Harmonic Gauge (GHG) formulation of GR, so our system is
symmetric-hyperbolic.

GHG has the practical advantage that we can prescribe free functions
called \textit{gauge source functions}, which appear in the choice of
gauge and determine the contracted Christoffel symbols. GHG also
implies the appearance of new constraints, called GHG constraints,
which can also be added to the equations of motion. Constraint
addition can be used for several purposes, for example to damp
constraint violations in the evolution of the
system~\cite{GunGarCal05}. Within the DF-GHG system the choice of
gauge source functions and constraint addition plays a crucial
role. It is by constraint addition that we get improved asymptotic
decay in one of the evolved fields, which permits $O(1)$ outgoing
lightspeed at~$\mathscr{I}^+$, a necessary condition for explicit time
integration schemes. Additionally, a suitable choice of gauge
regularizes the leading order decay of the metric fields for a big
class of initial data (ID), which is in turn necessary for the
rescaled evolved fields to be regular at~$\mathscr{I}^+$. In the
presence of non-vanishing physical radiation fields this
regularization procedure is performed through the introduction of a
\textit{gauge driver}, an additional evolved field whose value
at~$\mathscr{I}^+$ at each time is a function of the radiating field,
thus regularizing the field for which they are source
of~\cite{DuaFenGasHil22a}. This was explained and studied in full
three-dimensional simulations in model equations in~\cite{PetGauRai23}
and in a spherically symmetric reduction of the DF-GHG system
in~\cite{PetGauVan24}.

A common feature of both metric formulations that use hyperboloidal
slices is the appearance of formally singular terms. These are terms
comprised of a divergent factor multiplied by a field that decays
implicitly, so their product attains a finite limit
at~$\mathscr{I}^+$. These terms need to be carefully handled
numerically, because the factors appearing in them require explicit
evaluation for all grid-points that do not lie exactly
at~$\mathscr{I}^+$, evaluating ever bigger (smaller) terms as we
approach~$\mathscr{I}^+$. L'H\^opital's rule must moreover be
performed at the grid-points at~$\mathscr{I}^+$. Despite this,
successful implementations in spherical symmetry have been
reported. In our earlier work a free compactification
parameter~$n\in (1,2]$ was shown to give well defined properties at
the continuum level, but only a restricted subset~$n\sim(1.25,1.5]$
gave rise to smooth features close to~$\mathscr{I}^+$ so that our
simulations were found to converge with increasing resolution at the
expected rate. On the other hand, the DF-GHG system can only coincide
with conformal approaches for the exact value~$n=2$. Crucially,
various formally singular terms attain trivial limits for the
cases~$n<2$, but give~$O(1)$ contributions in the case~$n=2$. Our
earlier naive treatment restricted the range of values for the
parameter~$n$, which is the issue we overcame and report on in the
present work.

This paper addresses the following two questions. First, what
combination of constraint addition, gauge choice and definition of
reduction fields minimizes the number of formally singular terms
within the DF-GHG system? Second, with such a choice, can we adapt the
compactifying coordinate in order to get smooth, convergent numerical
simulations with the strongest permissible compactification
parameter,~$n=2$? We focus on spacetimes which already possess a BH in
the ID, which can be thought of as perturbations of the Schwarzschild
spacetime which eventually settle to another Schwarzschild BH. This
allows us consider the far-field asymptotics of the evolved fields
without the additional complication of managing a regular center. The
techniques studied here can be easily extended to the latter case as
in~\cite{PetGauVan24}.

The paper is organized as follows. In
section~\ref{Sec:Geometric_Setup} we explain the choice of the basic
metric fields and matter model, plus the formulation used to evolve
them. We then proceed in section~\ref{Sec:FST} to the details of the
regularization at~$\mathscr{I}^+$ at the continuum level, classify the
formally singular terms appearing in our previous work and how we can
get rid of some of them by a proper choice of reduction fields,
constraint addition and choice of gauge. We then go on in
section~\ref{Sec:Num_Ev} to present our numerical results for both
constraint violating and satisfying ID. We finalize by giving some
conclusions and outlook in section~\ref{Sec:Conclusions}.

\section{Geometric setup and the Einstein field equations}
\label{Sec:Geometric_Setup}

In this work we follow the same procedure as in our
earlier~\cite{PetGauVan24}, but for completeness nevertheless start
with an overview. We work in explicit spherical symmetry throughout
and consider asymptotically flat spacetimes, starting with spherical
polar coordinates~$X^{\underline{\mu}} = (T,R,\theta,\phi)$ in which
the metric asymptotes the standard Minkowski form in spherical polars
near spatial and future null infinity.

We take the first two evolved fields,~$C_+$ and~$C_-$, defined such
that
\begin{align}
  \xi^a = \p_T^a + C_+ \p_R^a\,,\qquad
  \underline{\xi}^a = \p_T^a+C_-\p_R^a
\end{align}
are null. The third evolved field, $\delta$, is defined through
\begin{align}
  \sigma_a = e^{-\delta}\xi_a\,,\qquad
  \underline{\sigma}_a = e^{-\delta}\underline{\xi}_a
\end{align}
satisfying the condition
\begin{align}
\sigma_a \p_R^a = -\underline{\sigma}_a \p_R^a = 1 \,.
\end{align}
Finally, the fourth metric variable, $\epsilon$, is defined through
the areal radius~$\mathring{R}$ as
\begin{align}
\mathring{R} \equiv e^{\epsilon/2} R \, .
\end{align}

In terms of~$\{ C_+, C_-, \delta, \epsilon \}$ the metric becomes
\begin{align}\label{Eq:Sph_metric_GHG}
(g_{\underline{\mu\nu}}) = \left(
\begin{array}{cccc}
 \frac{2 e^\delta C_+ C_-}{C_+ - C_-}
 & \frac{e^\delta \left(C_- + C_+ \right)}{C_- -C_+ }
 & 0 & 0 \\
 \frac{e^\delta \left(C_- +C_+ \right)}{C_- - C_+}
 & \frac{2 e^\delta}{C_+ - C_-}
 & 0 & 0 \\
 0 & 0 & \mathring{R}^2 & 0 \\
 0 & 0 & 0 & \mathring{R}^2 \sin^2 \theta \\
\end{array}
\right) \,.
\end{align}
The Levi-Civita derivative of~$g_{ab}$ is denoted by~$\nabla_a$. The
metric decomposes as
\begin{align}
g_{ab} = \mathbbm{g}_{ab} + \slashed{g}_{ab} \, ,
\end{align}
where~$\mathbbm{g}_{ab}$ is the~$\{T,R\}$ part and~$\slashed{g}_{ab}$
is the metric defined on the~$\{\theta,\phi \}$ sector.

Physics is governed by Einstein's equations, equivalent to
\begin{align}
  R_{ab} = 8 \pi \Big( T_{ab} - \frac{1}{2} g_{ab} T_c{}^c \Big) \,,
\label{trEFEs}
\end{align}
where~$R_{ab}$ is the Ricci tensor of~$g_{ab}$ and~$T_{ab}$ is the
stress-energy tensor. Following our symmetry assumptions, the
components of~$T_{ab}$ read
\begin{align}
( T_{\underline{\mu\nu}} ) = \left(
\begin{array}{cccc}
 T_{TT} & T_{TR} & 0 & 0 \\
 T_{TR} & T_{RR} & 0 & 0 \\
 0 & 0 & T_{\theta \theta} & 0 \\
 0 & 0 & 0 & T_{\theta \theta} \, \sin^2 \theta \\
\end{array}
\right) \, .
\end{align}
Due to spherical symmetry, all the metric fields and components of the
stress-energy tensor are functions of~$(T,R)$ only.

As a matter model we minimally couple a massless scalar field~$\psi$,
satisfying
\begin{align}
  \Box \psi \equiv g^{ab}\nabla_a\nabla_b\psi = 0 \, .
\end{align}
The stress energy tensor is then given by
\begin{align}\label{stressenergy_scalarfield}
  T_{\sigma\sigma} = (D_\sigma\psi)^2 \,,
  \quad T_{\underline{\sigma}\underline{\sigma}}
  = (D_{\underline{\sigma}}\psi)^2 \,,
  \non\\
  \quad T_{\sigma\underline{\sigma}} = 0 \,,
  \quad T_{\theta\theta} = \frac{e^\delta}{\kappa} \mathring{R}^2
  D_\sigma\psi D_{\underline{\sigma}}\psi\,,
\end{align}
where the subscripts~$\sigma$ and~$\underline{\sigma}$ denote
contraction with the null vectors~$\sigma^a$
and~$\underline{\sigma}^a$ respectively.

\subsection{Generalized Harmonic Gauge}
\label{Sec:GHG}

Generalized harmonic coordinates satisfy the
equations~$\Box X^{\ul{\alpha}}=F^{\ul{\alpha}}$,
where~$F^{\ul{\alpha}}$ are the gauge source functions, a set of free
functions that depend on the coordinates and metric components
only. This definition implies the constraints
\begin{align}
C^{\ul{\mu}} \equiv \Gamma^{\ul{\mu}} + F^{\ul{\mu}} =
0\,, \label{Eq:GHG_Constraint}
\end{align}
where~$\Gamma^{\ul{\mu}} = g^{\ul{\nu} \ul{\lambda}} \,
\Gamma^{\ul{\mu}}{}_{\ul{\nu} \ul{\lambda}}$ are the contracted
Christoffels with
\begin{align}
  \Gamma^{\underline{\mu}}{}_{\underline{\nu\lambda}}
  = \frac{1}{2} g^{\underline{\mu\rho}}
  (\p_{\ul{\nu}}g_{\underline{\rho\lambda}}
  + \p_{\ul{\lambda}} g_{\ul{\nu\rho}}
  - \p_{\underline{\rho}} g_{\ul{\nu\lambda}}) \, .
\end{align}
At the numerical level~$C^{\ul{\mu}}$ deviates from being identically
zero, so they measure the deviation from satisfying the GHG condition.

Using the GHG constraints we can define the reduced Einstein equations
(rEFEs),
\begin{align}
  R_{ab} - \nabla_{(a} C_{b)} + W_{ab} = 8\pi \Big(
  T_{ab} - \tfrac{1}{2} g_{ab} T_c{}^c \Big)\,,
  \label{eq:rEFEs_concise_form}
\end{align}
where~$W_{ab}=W_{(ab)}(C_c)$ is called the constraint addition tensor,
any symmetric rank-2 tensor which satisfies~$W_{ab}(0)=0$.  Curved
parentheses in subscripts denote the symmetrization in those
indices. Equation~\eqref{eq:rEFEs_concise_form} reduces
to~\eqref{trEFEs} when the constraints are satisfied, meaning that the
rEFEs and EFEs are equivalent for constraint satisfying solutions. The
rEFEs imply that metric components satisfy nonlinear curved-space wave
equations. Defining~$D_a$ as the covariant derivative associated
with~$\mathbbm{g}_{ab}$, they read
\begin{align}
  & D_\sigma\left(\frac{2}{\kappa}\mathring{R}^2D_{\ul{\sigma}}C_+ \right)
    +\mathring{R}D_\sigma\left(\mathring{R} F^\sigma \right)
    -D_\sigma\mathring{R}^2\frac{D_\sigma C_+}{\kappa}  \non \\
  & -\mathring{R}^2 \tilde{W}_{\sigma\sigma}
  = -8\pi\mathring{R}^2T_{\sigma\sigma} \,, \non \\
  & D_{\ul{\sigma}}\left(\frac{2}{\kappa}\mathring{R}^2D_\sigma C_-\right)
    -\mathring{R}D_{\ul{\sigma}}\left(\mathring{R}F^{\ul{\sigma}} \right)
    -D_{\ul{\sigma}}\mathring{R}^2\frac{D_{\ul{\sigma}} C_-}{\kappa}  \non \\
  & +\mathring{R}^2 \tilde{W}_{\ul{\sigma\sigma}}
  = 8\pi\mathring{R}^2T_{\ul{\sigma}\ul{\sigma}} \, , \non \\
  & {\Box}_2\delta + D_a(\mathbbm{g}^a{}_bF^b)
    + \frac{2e^\delta}{\kappa^3} \left[ D_{\ul{\sigma}}C_+ D_\sigma C_-
    - D_\sigma C_+ D_{\ul{\sigma}} C_- \right]  \non \\
  & +\frac{2}{\mathring{R}^2}\left(1-\frac{2M_{\mathrm{MS}}}{\mathring{R}}\right)
    +\frac{2e^\delta}{\kappa}\tilde{W}_{\sigma\ul{\sigma}}
    = \frac{16 \pi \, T_{\theta\theta}}{\mathring{R}^2} \, , \non \\
  & {\Box}_2\mathring{R}^2 - 2 -2R^2\tilde{W}_{\theta\theta}
    = -16\pi\frac{e^\delta}{\kappa}\mathring{R}^2 T_{\sigma\ul{\sigma}} \, .
    \label{Eqn:refes_metriceqs}
\end{align}
with
\begin{align}\label{Eqn:general_constrAdd}
  \tilde{W}_{\sigma\sigma} &= W_{\sigma\sigma} + \left( e^{-\delta}\p_R C_+
                             + D_\sigma\mathring{R} \right) C^\sigma \,, \non \\
  \tilde{W}_{\ul{\sigma\sigma}} &= W_{\ul{\sigma\sigma}}
                                  + \left( e^{-\delta}\p_R C_- + D_{\ul{\sigma}}
                                  \mathring{R} \right) C^{\ul{\sigma}} \,, \non \\
  \tilde{W}_{\sigma\ul{\sigma}} &= W_{\sigma\ul{\sigma}} \,,  \\
  \tilde{W}_{\theta\theta} &= W_{\theta\theta} + \mathring{R}\frac{e^\delta}{\kappa}
                             \left( D_{\ul{\sigma}}\mathring{R} \, C^\sigma
                             + D_\sigma\mathring{R}
                             \, C^{\ul{\sigma}} \right) R^{-2} \,\,, \non
\end{align}
where~${\Box}_2\equiv \mathbbm{g}^{ab}D_aD_b$ denotes the~$\{ TR \}$
d'Alembert operator and we define the
shorthand~$\kappa \equiv C_+ - C_-$. Finally, the Misner-Sharp
mass~\cite{MisSha64} is given by
\begin{align}
  M_{\mathrm{MS}}\equiv\tfrac{1}{2}\mathring{R}\bigg(
  2\frac{e^\delta}{\kappa} (D_\sigma \mathring{R}) (D_{\ul{\sigma}}
  \mathring{R}) + 1 \bigg)\,.\label{eqn:MS_defn}
\end{align}

Imposing the angular components of the GHG constraints implies
\begin{align}
  F^\theta = \mathring{R}^{-2} \cot \theta \, , \quad F^\phi = 0\,,
\end{align}
so the free components of the gauge source functions in spherical
symmetry are the null components~$F^\sigma$ and~$F^{\ul{\sigma}}$. In
terms of the free gauge sources and the chosen fields the null
components of the GHG constraints read
\begin{align}
  C^{\sigma} &\equiv C^a\sigma_a = F^\sigma
               +2\frac{D_{\ul{\sigma}}C_+}{\kappa}
               -2\frac{D_\sigma \mathring{R}}{\mathring{R}} \non \\
  C^{\ul{\sigma}} &\equiv C^a\ul{\sigma}_a = F^{\ul{\sigma}}
                    -2\frac{D_\sigma C_-}{\kappa}
                    -2\frac{D_{\ul{\sigma}}\mathring{R}}{\mathring{R}} \,.
                    \label{eq:GHG_constraints_GHG_coords}
\end{align}

Assuming a particular decay of the
constraints~\eqref{eq:GHG_constraints_GHG_coords} at~$\mathscr{I}^+$
allows one to read off the decay of incoming null derivatives of~$C_+$
and~$\epsilon$ after suitable choice of gauge source functions.  In
fact,~\cite{GasHil18} showed that it is possible to add multiples of
these constraints to the rEFEs so that these two incoming derivatives
decay as~$\mathring{R}^{-2}$, even when the constraints are
violated. In the next section we will use constraint addition for
achieving this task. Moreover, constraint addition will be used to
substitute formally singular terms by regular ones.

\section{Regularization at~$\mathscr{I}^+$}\label{Sec:FST}

The main point of the present work is to recast
equations~\eqref{Eqn:refes_metriceqs} in a form which is suited for
numerical implementation, with~$O(1)$ evolved fields
at~$\mathscr{I}^+$ and with the least formally singular terms
possible. This process involves choices of reduction fields,
rescalings, constraint addition and choice of gauge. Each of them will
be described in the following subsections, emphasizing which
conditions are necessary to reduce the number of formally singular
terms.

\subsection{\textit{Good-Bad-Ugly-F} classification}
\label{Sec:GoodBadUglyF}

Each of the metric fields in the DF-GHG system can be classified
according to its asymptotic decay with the \textit{Good-Bad-Ugly-F}
model, which corresponds to the following wave equations in Minkowski
spacetime
\begin{align} \label{Eq:GBUF}
 \Box g &= S_g \non \\
 \Box b &= (\p_T g)^2 + \frac{1}{R}\p_T f + S_b \non \\
 \Box u &= \frac{2}{R}\p_T u + S_u \non \\
 \Box f &= \frac{2}{R}\p_T f +2(\p_T g)^2 
\end{align}
where~$S_\varphi$ are called the sources of each equation, which are
assumed to decay at~$\mathscr{I}^+$ at least
as~$\mathring{R}^{-3}$. With the previous assumptions, it was proven
in~\cite{DuaFenGasHil22a} that, for suitable initial data, the
asymptotic decay of the~$g$ (good) and~$u$ (ugly) fields
towards~$\mathscr{I}^+$ is, respectively,
\begin{align} \label{Eq:gudecay}
 g &\sim \frac{G_1(T-R)}{R}  \\
 u &\sim \frac{m_u}{R} \non
\end{align}
where~$G_1$ is an~$O(1)$ function of retarded time and~$m_u$ is
constant for each cut of~$\mathscr{I}^+$. From the previous rates we
know that~$G\equiv Rg$ and~$U\equiv Ru$ are finite at~$\mathscr{I}^+$.

The reason for the inclusion of the~$f$ field in eq.~\eqref{Eq:GBUF}
is to regularize the leading order solution of the~$b$ (bad) field as
follows. Considering the system with~$f\equiv 0$, and with no equation
of motion for~$f$, called the \textit{Good-Bad-Ugly} model, the
leading order asymptotics of the~$b$ field is
\begin{align}
 b &\sim \frac{B_{1,1}(T-R) + B_{1,2}(T-R)\log R}{R} \,. \non
\end{align}
The logarithmic term in the previous expression is problematic when
one wants to evolve the rescaled~$b$ field so that it becomes~$O(1)$
at~$\mathscr{I}^+$. For model equations~\eqref{Eq:GBUF}, on the other
hand,
\begin{align}
 b &\sim \frac{B_1(T-R)}{R} \,, \non
\end{align}
essentially having the same type of asymptotics as the~$g$
field. Importantly,~$f$ has the same leading order asymptotics, with
its corresponding leading order satisfying
\begin{equation}
 F_1'(T-R) = -G_1'(T-R)^2/2 \,,
 \label{Eq:Fasymptotics}
\end{equation}
thus targeting the source for the bad asymptotic behavior of the~$b$
field, effectively regularizing its leading order at~$\mathscr{I}^+$.

The asymptotic decay~\eqref{Eq:gudecay} also implies that outgoing
null derivatives of~$g$ and~$u$ decay as
\begin{align}
  D_\sigma g \sim D_\sigma u \sim \mathring{R}^{-2}\,.
\end{align}
Incoming null derivatives, on the other hand, decay as
\begin{align}
 D_{\ul{\sigma}} g &\sim \mathring{R}^{-1} \non \\
 D_{\ul{\sigma}} u &\sim \mathring{R}^{-2} \,. \non
\end{align}
The~$\mathring{R}^{-2}$ decay of this null derivative of the ugly
field is what has been mentioned previously as improved decay.

\subsection{Compactified hyperboloidal foliations}
\label{Sec:Hyperboloidal}

Formally singular terms arise by transforming
equations~\eqref{Eqn:refes_metriceqs} to compactified hyperboloidal
coordinates, as compactification introduces divergences in the
equations which need to be compensated by appropriate decay of the
evolved fields. We therefore start by describing the procedure to
foliate spacetime with compactified hyperboloidal slices.

We introduce a compactified radial coordinate $r$ according to
\begin{align}
  R(r) &= \frac{r}{\Omega(r)^{\frac{1}{n-1}}}
         \, , \non \\
  \Omega(r)& =1-\frac{r^2}{r_{\mathscr{I}}^2} \, ,
             \quad 1< n \leq 2 \, ,
\end{align}
so that~$R\rightarrow \infty$ is mapped
to~$r\rightarrow r_{\mathscr{I}}$, which is taken to be finite. The
derivative of~$R(r)$ grows asymptotically as~$R'\sim R^n$, meaning
that the free parameter~$n$ parametrizes the rate at which the
original radial coordinate~$R$ is compactified~\cite{CalGunHil05}.

Using a height function~$H(R)$, we next define a new time function so
that constant time slices are lifted from the Cauchy ones to
intersect~$\mathscr{I}^+$. Specifically we take
\begin{align}
  t = T - H(R) \,.
  \label{eq:Height_func}
\end{align}
In order for constant~$t$ slices to reach~$\mathscr{I}^+$, the time
function~$t$ needs to asymptote to retarded time. We choose
\begin{align}
  H(R) = R - m_{C_+}\ln R - r \,.
  \label{Eqn:heightfunction_withmass}
\end{align}
The requirement that the variable~$C_+$ has improved asymptotic decay
can be seen clearly from
expression~\eqref{Eqn:heightfunction_withmass}, since the
term~$m_{C_+}$ must be constant.

The Jacobians from Cauchy~$(T,R)$ to hyperboloidal coordinates~$(t,r)$
are
\begin{align}\label{Eq:HFjacobians}
 &\p_R  = \frac{1}{R'(r)}\p_r
  + \left( 1 - \frac{1}{R'(r)}
  -\frac{ m_{C_+} }{ R(r) } \right) \p_t  \, , \non \\
 &\p_T = \p_t \,.
\end{align}

With the Jacobians at hand we could transform the metric~$g_{ab}$ to
the hyperboloidal basis, after which it would be natural to ask
whether the rate at which~$g_{ab}$ diverges when
approaching~$\mathscr{I}^+$ admits a conformal compactification, that
is, demanding that a conformally rescaled metric,
\begin{align}
 \tilde{g}_{ab} = \Xi^2 g_{ab} \,,
\end{align}
be regular at~$\mathscr{I}^+$. Assuming that~$\Xi$ decays
asymptotically as~$\sim R^{-1}$, the metric~$\tilde{g}_{ab}$ can only
be regular if the leading divergence of~$g_{ab}$ goes at the
rate~$\sim R^2$. From~\eqref{Eq:HFjacobians} we see that the Cauchy to
hyperboloidal Jacobians depend on~$R'(r)$, so the metric~$g_{ab}$ in
hyperboloidal coordinates will only do so for the choice of
parameter~$n=2$ (for further details see~\cite{HilHarBug16}).  So
motivated we choose~$n=2$ throughout this work. This is our main
result to report, and will allow us to directly compare our results to
those obtained with conformal compactification. On the other hand,
from~\cite{PetGauVan24}, we know that this choice of~$n$ is not
essential for the inclusion of~$\mathscr{I}^+$ in numerical
simulations of the DF-GHG system.

The case~$n=2$ leads to an interesting subtlety as opposed
to~$n<2$. For the latter, the outgoing coordinate lightspeed~$c_+$
tends to one when approaching~$\mathscr{I}^+$. For the~$n=2$
case,~$c_+$ at~$\mathscr{I}^+$ depends on the value chosen
for~$r_{\mathscr{I}}$, and has a flatter profile that approaches one
for larger~$r_{\mathscr{I}}$. We thus choose the
value~$r_{\mathscr{I}} = 20$ in all cases reported here.

\subsection{First order reduction and rescaling}
\label{Sec:FOR}

The purpose of using compactified hyperboloidal coordinates is to
extract directly radiation terms at~$\mathscr{I}^+$ from numerical
simulations. Therefore, for the evolved fields to be~$O(1)$
at~$\mathscr{I}^+$ we rescale them as~$$Z \equiv R\zeta\,,$$
where~$\zeta$ stands for
either~$\{\delta,\, \epsilon,\, \psi,\, f_{D} \}$. Similarly we
take $$\tilde{C}_- \equiv R(C_-+1)\,,$$ where~$C_-=-1$ corresponds to
its Minkowski value, so it is only this difference that decays, and
the \textit{Good-Bad-Ugly-F} classification informs us how radial null
derivatives decay. We perform a full first order reduction using these
null derivatives, and rescale them as
\begin{align}\label{Eq:redvars}
 Z^+ &\equiv R^2D_\sigma\zeta \,, \quad
 Z^- \equiv R\,D_{\ul{\sigma}}\zeta \,, 
\end{align}
and, analogously,
\begin{align}\label{Eq:redvarCm}
 \Theta^- &\equiv R^2 \frac{D_\sigma C_-}{\kappa} \,,\quad
 \ul{\Theta}^- \equiv R\,\frac{D_{\ul{\sigma}}C_-}{\kappa} \,.
\end{align}
Definitions~\eqref{Eq:redvars}-\eqref{Eq:redvarCm} imply reduction
constraints, which must be met in order for the first order system to
be equivalent to the original second order one. The reduction
constraints read
\begin{align}\label{Eq:FORconstraints}
  \p_R Z -\frac{1}{R}Z +\frac{ e^{\Delta /R} }{\kappa}
  \left( Z^- -\frac{1}{R}Z^+ \right) = 0 \,, \non \\
  \p_R \tilde{C}_- -\frac{1}{R}\tilde{C}_-
  + e^{\Delta /R} \left( \ul{\Theta}^- -\frac{1}{R}\Theta^- \right) = 0 \,.
\end{align}
To examine these constraints in hyperboloidal coordinates the $R$
derivative should be replaced using~\ref{Eq:HFjacobians} and the $t$
derivative with the reduction fields definition, so only spatial
derivatives appear. For compactness we prefer the
form~\ref{Eq:FORconstraints}.

Recalling that incoming null derivatives of ugly fields have improved
asymptotic decay, one could be tempted to rescale
their~$D_{\ul{\sigma}}$ derivative by an extra power of~$R$. This,
however, can only be done if~$S_u\sim O(\mathring{R}^{-4})$, as was
done in~\cite{GasGauHil19}. Slower decay of~$S_u$ leads to potentially
singular terms.

In what follows we will take a choice of gauge and constraint addition
so that~$C_+$ and~$\epsilon$ satisfy ugly wave
equations. Importantly,~$S_\epsilon \sim\mathring{R}^{-3}$, so we keep
definitions~\eqref{Eq:redvars} for this variable, eventually leading
to one formally singular term in the evolution equation for
$E^-\equiv R\,D_{\ul{\sigma}}\epsilon$\,.  On the other
hand,~$S_{C_+}\sim \mathring{R}^{-4}$, so for this one special case we
take as reduction fields for~$C_+$ the variables
\begin{align}\label{Eq:redCp}
 \Theta^+ &\equiv R^2 \frac{D\sigma C_+}{\kappa} \\
 \ul{\Theta}^+ &\equiv R^2 \frac{D_{\ul{\sigma}} C_+}{\kappa} \non
\end{align}
and the equation of motion for these variables are fully regular. This
is in contrast to our previous work, where the incoming null
derivative of~$C_+$ was rescaled only by a single power of~$R$, the
evolved variable had additional decay, and its equation of motion
contained formally singular terms.

One last comment about rescaling is yet to be made. The~$\{T,R\}$ wave
operator is built upon successive outgoing-ingoing or ingoing-outgoing
radial null derivatives. This implies that the equation of motion
for~$Z^-$ has a~$D_\sigma$ derivative. When solving for~$\p_\tau Z^-$
we get the following~$O(1)$ factor coming from the Jacobians
\begin{align}
 \p_\tau Z^- = \frac{1}{R'(1-H'C_+)}[...]
 \label{Eq:outgoingJacobian}
\end{align}
where~$H'=dH/dR$. The choice~$C_+\equiv 1+\tilde{C}_+/R$ makes the
denominator in eq.~\eqref{Eq:outgoingJacobian} become
\begin{align*}
  R'(1-H'C_+) = 1 + \frac{R'}{R}(m_{C_+}-\tilde{C}_+)
  + \frac{\tilde{C}_+}{R} +\frac{R'}{R^2}\tilde{C}_+m_{C_+}
\end{align*}
for which the second term on the right hand side is formally
singular. Therefore, we rather choose the~$C_+$ rescaling as
\begin{align}\label{Eq:CpRescaling}
 C_+ \equiv 1 + \frac{m_{C_+}}{R} + \frac{\hat{C}_+}{R^2} \,.
\end{align}
With this definition the Jacobian factor~\eqref{Eq:outgoingJacobian}
is explicitly regular. The heuristic results
of~\cite{DuaFenGasHil22a}, however, do not guarantee that~$\hat{C}_+$
is bounded during evolution. This rescaling should be interpreted as a
test of whether~$\hat{C}_+$ remains bounded in the time development or
not. The crucial reason for this Jacobian to be special is that the
height function is built from the~$C_+$ variable, which in turn
appears in the~$D_\sigma$ derivative. With definitions~\ref{Eq:redCp}
and~\ref{Eq:CpRescaling} the reduction constraint for these variables
reads
\begin{align}\label{Eq:FORconstraintCp}
  \p_R \hat{C}_+ -m_{C_+} -\frac{2}{R}\hat{C}_+
  + e^{\Delta /R} \left( \ul{\Theta}^+ -\Theta^+ \right) = 0 \,.
\end{align}

\subsection{Gauge source functions}
\label{Sec:GSFs}

A suitable choice of gauge is needed for the rescaled evolved fields
to be~$O(1)$ at~$\mathscr{I}^+$. Following~\cite{DuaFenGasHil22a} the
leading order term has a fixed expression, so we can write the null
components of the gauge source functions as
\begin{align}
  F^\sigma &= \frac{2}{\mathring{R}}
             + \frac{1}{\mathring{R}^2}\bar{F}^\sigma \\
  F^{\ul{\sigma}} &= -\frac{2}{\mathring{R}}
                    + \frac{1}{\mathring{R}^2}\bar{F}^{\ul{\sigma}}
                    \,. \non
\end{align}
The similarity of this choice and the \textit{scri-fixing} choice
of~\cite{Zen08} is noteworthy. The next to leading order term is
given by
\begin{align}
  \bar{F}^\sigma &= e^{-\epsilon/2}\left( m_{C_+}
                   + \frac{\hat{C}_+}{R} \right) \\
 \bar{F}^{\ul{\sigma}} &= e^{-\epsilon/2} F_D \,. \non
\end{align}
The first term in~$\bar{F}^\sigma$ is needed for the~$\Theta^\pm$
evolution equations to be regular and for Schwarzschild spacetime in
Kerr-Schild coordinates to be a static solution of the equations of
motion. For the~$\bar{F}^{\ul{\sigma}}$ component, the
field~$F_D \equiv Rf_D$ is a rescaling of the \textit{gauge driver}
field~$f_D$. This is needed to regularize the~$C_-$ field in the
presence of radiation, as discussed in~\cite{DuaFenGasHil22a}. In
spherical vacuum we take~$f_D\equiv 0$, but in the presence of a
radiating field, such as the massless scalar field we use here, it
satisfies the equation
\begin{equation}
 \Box f_D -\frac{2}{R}\p_Tf_D -32\pi\left(\p_T\psi\right)^2 = 0 \,.
 \label{Eq:gaugedriver}
\end{equation}

Thus~$f_D$ plays the role of the~$f$ field from the GBUF model, and
regularizes the bad asymptotics of the~$C_-$ field in pure harmonic
gauge. We rescale the null derivatives of~$f_D$ according to the same
recipe as in equation~\eqref{Eq:redvars}. From~\eqref{Eq:gaugedriver}
and the analogy to the~$f$ field in the GBUF model,
the~$F_{D}^-\equiv RD_{\ul{\sigma}}f_D$ will have a formally singular
term in its equation of motion.

\subsection{Constraint addition}
\label{Sec:Cadd}

There are two crucial parts of constraint addition for the present
setup to work. The first, identical to our previous approach, is to
add constraints to the~$C_+$ and~$\epsilon$ equations of motion in
order to make these fields satisfy wave equations of `ugly' type in
our classification. The possibility for doing so relies on the fact
that the constraints~\eqref{eq:GHG_constraints_GHG_coords} have the
exact terms which can turn a good wave equation into an ugly one for
these two fields. This part of constraint addition reads
\begin{align}
  W_{\sigma\sigma} &= -\left( e^{-\delta}\p_R C_+
                     + D_\sigma\mathring{R} \right) C^\sigma
                     \,, \nonumber\\
  W_{\theta\theta} &= -\mathring{R}R^{-2}\frac{e^\delta}{\kappa}
                     \left( D_{\ul{\sigma}}\mathring{R} \,C^\sigma
                     + D_\sigma\mathring{R} \,C^{\ul{\sigma}} \right)
                     \,.\label{Eq:ConstAddUglify}
\end{align}
The choices~\eqref{Eq:ConstAddUglify}
render~$\tilde{W}_{\sigma\sigma}=\tilde{W}_{\theta\theta}=0$, making
the~$C_+$ and~$\epsilon$ equations of motion more compact.

The second part of constraint addition is useful for avoiding formally
singular terms. Contrary to the GBUF model in Minkowski spacetime,
where the only formally singular terms arise in the evolution equation
for rescaled incoming null derivative of the ugly field and gauge
driver, the rEFEs with solely the previous constraint addition possess
another type: terms proportional to~$E^-$ in the evolution equations
for~$\ul{\Theta}^-$ and~$\Delta^-$. A formally singular term of the
same type appears when one computes~$R^2C^{\ul{\sigma}}$ in terms of
our reduction fields, due to the fact that~$D_{\ul{\sigma}}\epsilon$
is only rescaled by a single power or~$R$. Note that, even when
constraints are violated, the combination~$R^2C^{\ul{\sigma}}$
is~$O(1)$ by virtue of~$\epsilon$ having improved asymptotic decay,
meaning that this technique only works because~$\epsilon$ decays
better. The particular constraint addition we use which improves
regularity of the~$\Delta^-$ and~$\ul{\Theta}^-$ equations is
\begin{align}
  W_{\sigma\ul{\sigma}} &= \frac{1}{2\mathring{R}}
                          \left( C^\sigma - C^{\ul{\sigma}} \right) \,, \non\\
  W_{\ul{\sigma\sigma}} &= - D_{\ul{\sigma}}\mathring{R} \, C^{\ul{\sigma}}
                          \,. \label{Eq:ConsAddRegularize}
\end{align}
When solving for~$\p_\tau\ul{\Theta}^-$ and~$\p_\tau\Delta^+$ in the
equations of motion the previous constraint
addition~\eqref{Eq:ConsAddRegularize} corresponds to adding a multiple
of the constraint~$R^2C^{\ul{\sigma}}$, effectively removing the
formally singular terms previously mentioned. In summary, the total
constraint addition tensor we include corresponds to the different
components appearing in~\eqref{Eq:ConstAddUglify}
and~\eqref{Eq:ConsAddRegularize}.

\section{Numerical Evolutions}\label{Sec:Num_Ev}

Our purpose is to test whether the removal of formally singular terms
allows well-behaved numerics for the strongest permissible
compactification parameter,~$n=2$. For simplicity we restrict to
spacetimes already containing a black hole in the initial data, and so
avoid dealing with the regular center. We excise the black hole
region. Since the excision procedure requires the use of regular
coordinates across the horizon, we take initial data as perturbations
of the Schwarzschild solution written in Kerr-Schild coordinates,
which are horizon-penetrating. In terms of our variables this solution
reads
\begin{align}
  & C_+ \equiv \frac{1-\frac{2M}{R}}{1+\frac{2M}{R}} \, ,
    \quad C_- \equiv -1  \, , \quad \delta \equiv 0\,,
    \quad  \epsilon \equiv 0 \,.
\label{Eqn:SS_sol}
\end{align}
Equation~\eqref{Eqn:SS_sol} is an exact solution of the rEFEs with our
choice of gauge source functions. Therefore, it is seen as a static
solution in our numerical implementation up to truncation error. Our
units are set by the choice~$M=1$, which we take in all our
simulations. Finally, from expression~\ref{Eqn:SS_sol} we get
$m_{C_+} = -4M$, which is the value we feed to the height function.

For the numerical implementation we use standard techniques. The
evolution consists of a method of lines, evolving in time with a
fourth-order Runge-Kutta scheme. We discretize the spatial slices by
the use of a non-staggered grid, with equally spaced grid-points in
the~$r$ coordinate, with a base resolution of 200 grid-points in all
cases, where the first and last grid-points lie exactly at the
excision radius and~$\mathscr{I}^+$, respectively, corresponding
to~$r\in [1.6, 20]$\,.  As spatial discretization we use second-order
accurate centered finite differences except for the grid-point
at~$\mathscr{I}^+$, for which we use a one-sided stencil towards the
inner domain following a truncation error matching
technique~\cite{GauVanHil21}. At this grid-point we evaluate the
remaining formally singular terms using l'H\^opital's rule, the
standard technique used in all previous implementations of
hyperboloidal coordinates in NR. In GHG there are no superluminal
speeds, making both the event horizon and~$\mathscr{I}^+$ purely
outflow boundaries. There is therefore no need for boundary conditions
in this formulation, so it suffices to extrapolate the numerical data
to populate ghost points beyond~$r_{\mathscr{I}}$ or inside the
excised region. We use fourth-order polynomial extrapolation for this
task. Ghost points are used to compute Kreiss-Oliger
dissipation~\cite{KreOli73}, which is added to the evolution equation
to improve convergence, with a dissipation parameter of~$0.02$.

\subsection{Constraint violating Schwarzschild perturbations}
\label{SubSec:constraintviolating}

\begin{figure*}[t]
  \includegraphics[scale=0.38]{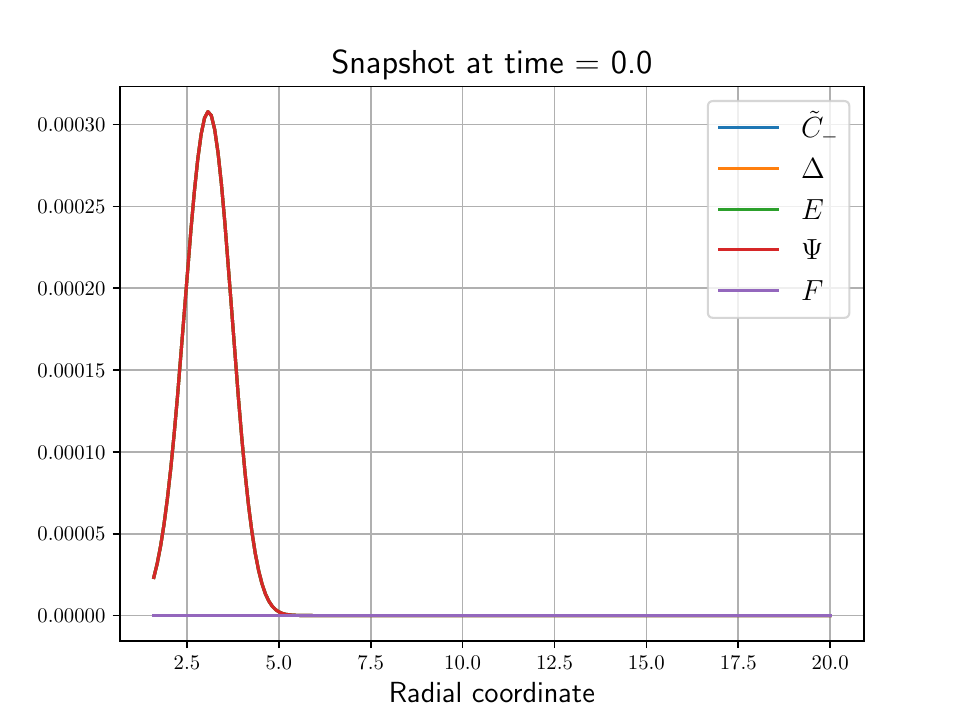}
  \hspace{-0.5cm}\includegraphics[scale=0.38]{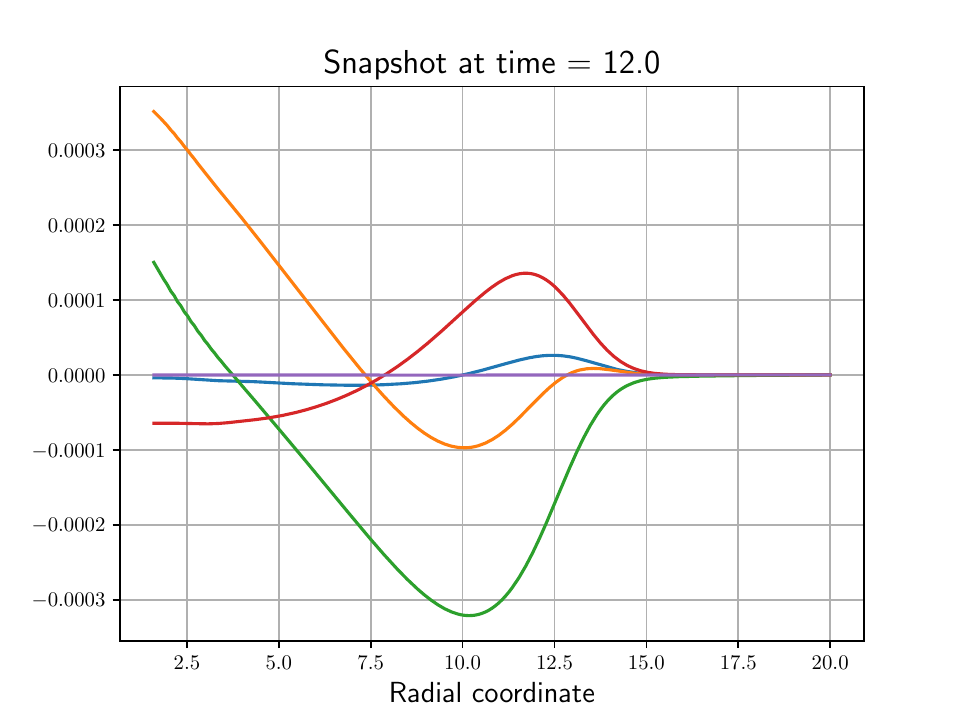}
  \hspace{-0.5cm}\includegraphics[scale=0.38]{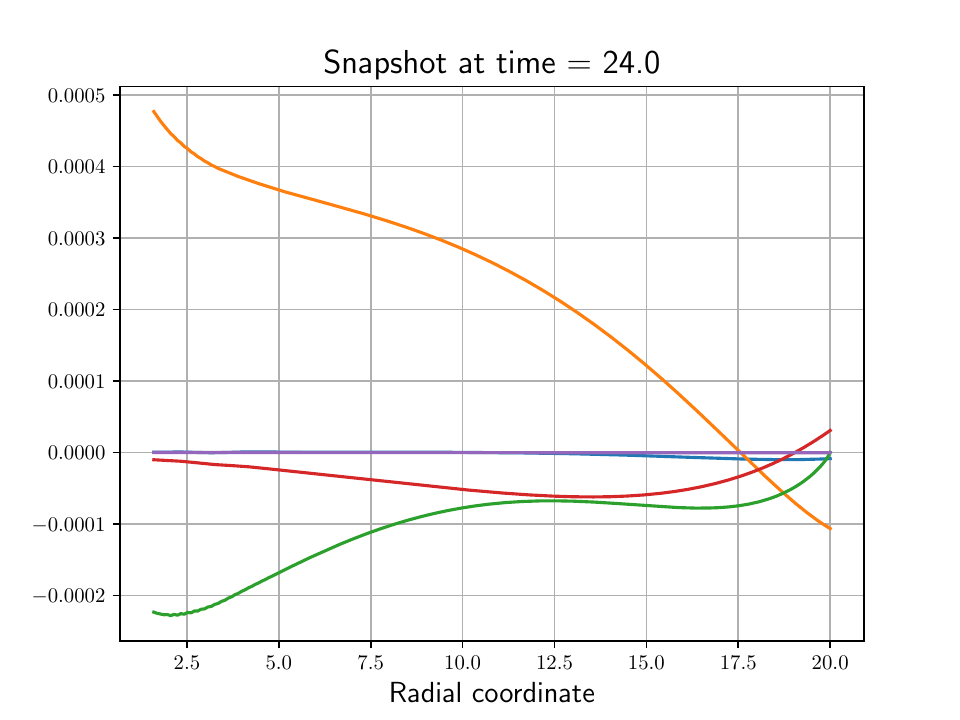}
  \caption{In these plots we show snapshots of the evolved
    variables~$\hat{C}_-$, $\Delta$, $E$, $\Psi$ and~$F_D$ from our
    numerical evolutions.  Observe that, as expected, the field~$E$
    vanish at~$\mathscr{I}^+$ ($r=20$ in our coordinates) at all times
    whereas the others oscillate. At later times the remnant features
    continue to shrink.\label{Fig:cviolating_snapshots} }
\end{figure*}

As mentioned above, the DF-GHG system is an evolution system with
constraints. Efficient numerical simulations for applications in GW
astronomy, however, demand for the use of free evolution schemes, for
which we do not solve the constraints at each timestep. Since
constraint violations are always present due to truncation error, we
need to test our evolution scheme for constraint violating
data. Therefore, as a first test we start with ID that is Hamiltonian,
Momentum and GHG constraint violating. We take a Gaussian pulse
centered at~$R=3M$ added on top of the Schwarzschild solution, with
zero time derivatives. This ID corresponds to taking
\begin{align}
& C_+(0,r) = \frac{1-\frac{2}{R}}{1+\frac{2}{R}}
  + C_{p_0} e^{-(R-3)^2} \, , \non \\
& C_-(0,r) = -1 + C_{m_0} e^{-(R-3)^2} \, , \non \\
& \delta(0,r) = \delta_0 e^{-(R-3)^2} \, ,
  \quad \epsilon(0,r) = \epsilon_0 e^{-(R-3)^2} \, , \non \\
& \psi(0,r) = \psi_0 e^{-(R-3)^2} \, ,
  \quad f_{D}(0,r) = 0 \,, \non \label{Eqn:SS_Constraint_Violating_ID}
\end{align}
with the rescaled reduction fields computed accordingly from their
definitions. Because of this last fact, reduction constraints are
satisfied at the ID, and their violation arise during evolution just
because of truncation error.

For this test we considered small perturbations so that the apparent
horizon does not grow significantly during evolution. In particular,
we chose the initial amplitude of the fields to
be~$C_{p_0} = C_{m_0} = \delta_0 = \epsilon_0 = \psi_0 = 10^{-4}$.
Snapshots of the numerical evolution with the previous ID are shown in
figure~\ref{Fig:cviolating_snapshots}.

As expected, the rescaled gauge driver field~$F_D$ departs from zero
from the first timesteps, targeting the correct asymptotic value given
by the analogous of expression to~\eqref{Eq:Fasymptotics}. The square
power in that expression makes the plot of this field invisible in the
scale of the other fields.~$F_D$'s presence, however, regularizes the
evolution of the~$C_-$ field, as can bee seen from
figure~\ref{Fig:cviolating_snapshots}, displaying a smooth behavior
at~$\mathscr{I}^+$ for all times.

\begin{figure}[t!]
 \includegraphics[scale=0.4]{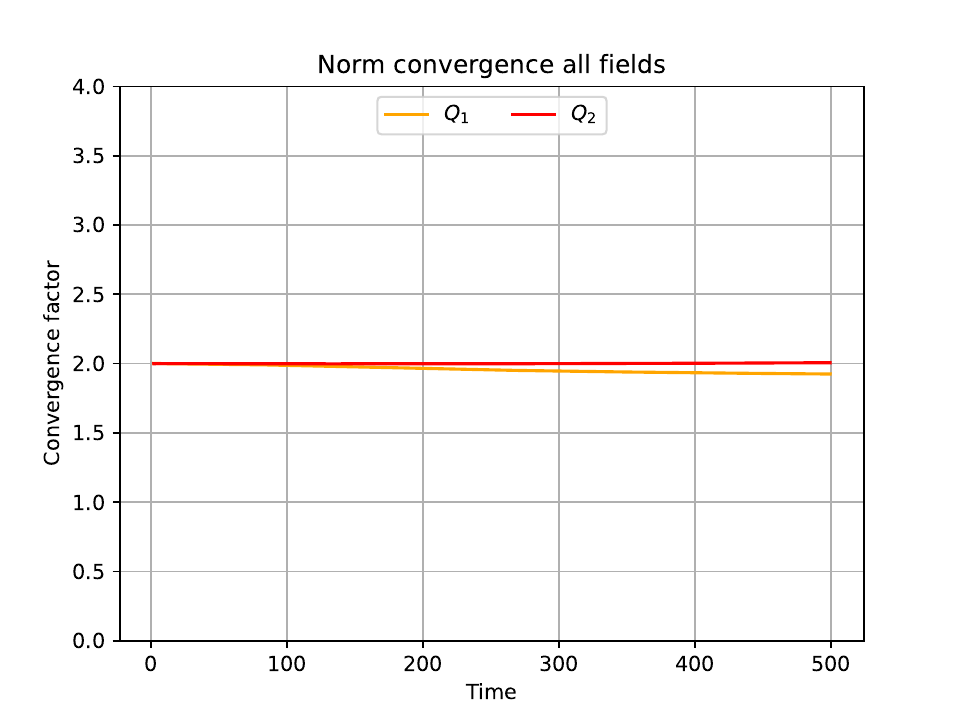}
 \caption{In this figure we plot the norm self-convergence rate for
   evolutions of constraint violating data on top of a Schwarzschild
   black hole. The two curves correspond to the rate obtained with
   three resolutions, being~$Q_1$ obtained with the lowest three
   and~$Q_2$ with the highest. Our second order accurate spatial
   finite differences imply that we should expect this curves to lie
   at two. Although not lying perfectly at two, the curves improve as
   we increase resolution (~$Q_2$ compared to~$Q_1$). Details of the
   initial data and norms are given in the main text.}
 \label{norm_convergence}
\end{figure}

We proceeded to increase resolution by doubling the number of
grid-points three times, starting with 200 grid-points. We then
computed the norm of the difference of the very high (V) and high (H)
resolutions, the high and medium (M) resolutions, and the medium and
low (L) resolutions, with all these differences computed at the low
resolution grid-points, by use of the norm
\begin{align}
  \int \left[ r^2Z^2 + \left( \frac{2R'-1}{2R^2}(Z^+)^2
  + \frac{1}{2}(Z^-)^2 \right)  \right]dr
\end{align}
We then computed the convergence factors
\begin{align}
  Q_1&=\log_2\left(\frac{||M-L||}{||H-M||}\right)\,,
       \quad Q_2=\log_2\left(\frac{||H-M||}{||V-H||}\right)\,, \non
\end{align}
as a function of time. The result of the previous computation with our
numerical data is shown in figure~\ref{norm_convergence}. The use of
second-order accurate finite differences implies that the convergence
factor must tend to the number 2 as we increase resolution. Looking at
figure~\ref{norm_convergence} we note that the numerical error of our
evolution decreases at the expected rate for increasing resolution, in
favor of well-behaved numerics tending to the continuum solution in
the limit of infinite resolution.

\subsection{Constraint satisfying Schwarzschild perturbations}
\label{SubSec:GHGID_numerics}

Solutions of the rEFEs are only equivalent to those of the EFEs when
the constraints are satisfied. We therefore proceeded to the case of
most physical interest, which corresponds to ID that satisfies all the
constraints. To solve for the ID we follow the same technique as in
our previous work. Here we will describe the procedure only
briefly. For the interested reader we suggest
following~\cite{PetGauVan24}.

The essential features for finding constraint satisfyingly ID with our
method go as follows. We start by taking the exact Schwarzschild
values for the variables~$C_\pm (0,r)$, $\epsilon (0,r)$
and~${^\gamma K_{rr}}\equiv\gamma^{rr}K_{rr}$, where~$\gamma^{rr}$
and~$K_{rr}$ correspond to the~$rr$ components of the inverse spatial
metric and extrinsic curvature in hyperboloidal coordinates,
respectively. This last choice determines~$\p_\tau \delta (0,r)$.  We
solve the Hamiltonian and Momentum constraints to
determine~$\delta (0,r)$ and~$\p_\tau\epsilon (0,r)$, under the
assumptions that~$\psi(0,r) = \psi_0 e^{-(R-3)^2}$
and~$n^a\nabla_a\psi=0$, where~$n^a$ is the unit normal to the spatial
hypersurfaces of our foliation and we take~$\psi_0=10^{-4}$. Finally,
GHG constraints are used to solve for~$\p_\tau C_\pm (0,r)
\,$. Exploiting the freedom in the choice of ID for~$f_D$, we
choose~$f_D (0,r)$ such that~$\p_\tau C_- (0,r)=0$
with~$\p_\tau f_D (0,r)=0\,$.

\begin{figure}[t!]
 \includegraphics[scale=0.4]{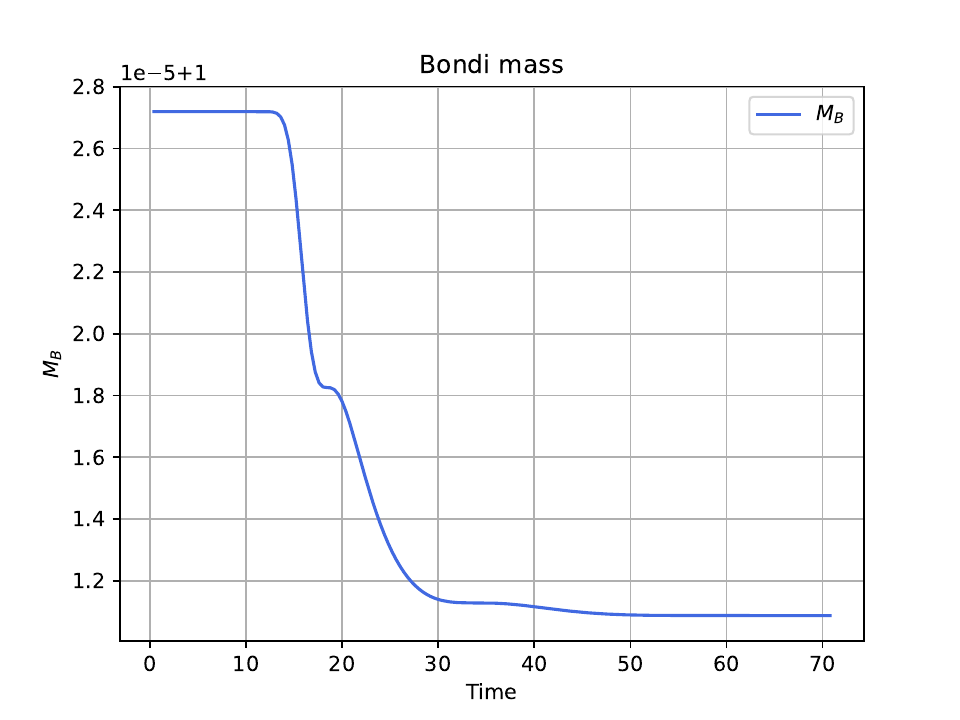}
 \caption{In this figure we plot the Bondi mass as a function
   of~$\tau$. The expected properties for physically-viable solutions
   are verified: it is monotonically decreasing, and most of its
   decrease happens as radiation leaves the domain
   through~$\mathscr{I}^+$, and settles asymptotically to a constant
   value slightly above the original mass of the perturbed black
   hole.}
 \label{Bondimassfig}
\end{figure}

The evolution is qualitatively similar to that of constraint violating
data. Regular behavior at~$\mathscr{I}^+$ is seen in all variables, so
we omit snapshots. Constraint violations arise in the time evolution
of the system, as expected. They remain at least an order of magnitude
smaller than the equivalent quantities in the constraint violating
data even at our base resolution, thus closer to the correct physical
modeling of the system at hand. Constraint violations could be
improved by constraint damping terms incorporated into~$W_{ab}$. These
are, however, incompatible with our constraint addition terms for
regularizing the asymptotics in the wavezone and avoiding formally
singular terms, so we decided not to add them here. The use of a
smooth transition from constraint damping terms to our choices
at~$\mathscr{I}^+$ is left for future work.

\begin{figure}[t!]
 \includegraphics[scale=0.4]{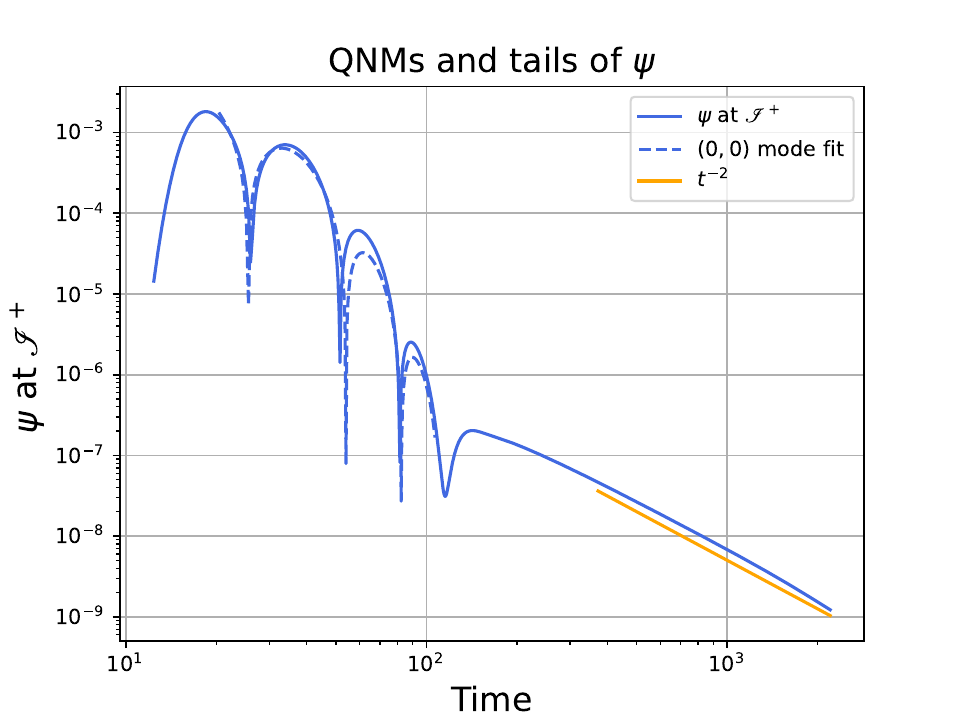}
 \caption{In this plot we show the rescaled scalar field~$\Psi$ at
   future null infinity as a function of hyperboloidal time for
   constrained-solved initial data (details in main text). The
   amplitude is sufficiently small so the nonlinear evolution
   resembles the Cowling approximation of a scalar field on a
   Schwarzschild background. In the first part of the plot we fit the
   fundamental spherical QNM~\cite{BerCarSta09} with complex
   frequency~$\omega M = 0.11 + 0.10i$. For the second part the field
   decays as a~$t^{-2}$ power law, known as the Price tail.}
 \label{QNMs_tails}
\end{figure}

To compare with known physics we extract the value of the variables
at~$\mathscr{I}^+$ as functions of time. We plot the Bondi mass,~$M_B$
in figure~\ref{Bondimassfig}. It is positive, monotonically decreasing
and settles to a constant value. The late-time value to which~$M_B$
settles is slightly above~$M$, consistent with a BH that accretes part
of the scalar field, the rest being radiated
through~$\mathscr{I}^+$. We also plot the value of~$\Psi$
at~$\mathscr{I}^+$ in figure~\ref{QNMs_tails}. As the perturbation on
top of Schwarzschild spacetime is small, we recover the expected
behavior for linear perturbations. This corresponds to the spherically
symmetric quasi-normal mode of the scalar field for the first part of
the data and~$t^{-2}$ decay at late times~\cite{Pri72}.

\section{Conclusions}\label{Sec:Conclusions}

In this paper we have demonstrated the use of an aggressive
compactification in nonlinear numerical evolutions that
include~$\mathscr{I}^+$. We made use of GHG with the DF technique, for
which both a Cauchy and a compactified hyperboloidal slicing of
spacetime are assumed, allowing for a symmetric-hyperbolic set of
evolution equations and the inclusion of~$\mathscr{I}^+$
simultaneously. We restricted to spherical symmetry, where the
computation and understanding of the equations of motion is simpler,
but which has all the fundamental difficulties of regularization and
formally singular terms present in the full 3d case.

We performed a detailed study of how to remove a subset of formally
singular terms from the equations by a different constraint addition
and choice of reduction variables as compared to our previous
work. This allowed us to reach self-convergent numerics for the
strongest compactification parameter,~$n=2$. The use of this specific
parameter is important to compare with numerical results using a
conformal compactification, the idea that follows more closely
Penrose's proposal on defining asymptotics of spacetime.

Numerical evolutions where performed for both constraint violating and
satisfying data, with both corresponding to small deviations of a
Schwarzschild BH. In the constraint satisfying case, we found strong
evidence that our numerics capture well-established physics, getting
the expected behavior of~$M_B$ at~$\mathscr{I}^+$, as well as
recovering the spherically-symmetric quasinormal mode and the tail
decay rate for the scalar field at~$\mathscr{I}^+$, both predicted by
linear theory.

Having successfully managed the spherical case over the complete range
of compactification parameters, our focus now is to extend the
approach to full 3d. We will report on our progress on this front
elsewhere.

\acknowledgements

The authors thank Shalabh Gautam and Alex Va\~n\'o-Vi\~nuales for
helpful discussions and comments on the manuscript. The Mathematica
notebooks associated with this work can be found
at~\cite{PetHil25_zenodo_web}. This work was supported through FCT
(Portugal) Project No. UIDB/00099/2020, by PeX-FCT (Portugal) program
2023.12549.PEX, by funding with DOI 10.54499/DL57/2016/CP1384/CT0090,
as well as IST-ID through Project No. 1801P.00970.1.01.01.

\normalem
\bibliography{SphGR_massfields_DFGHG.bbl}

\end{document}